\documentclass[prx,twocolumn,showpacs,amsmath,amssymb,superscriptaddress]{revtex4-2}
\usepackage{graphicx}
\usepackage{dcolumn}
\usepackage{bm}
\usepackage{hyperref}
\hypersetup{colorlinks,citecolor=blue, filecolor=blue, linkcolor=blue , urlcolor=blue}

\begin{document}

\title{In-plane magnetocrystalline anisotropy in the van der Waals antiferromagnet \\ FePSe$_3$ probed by magneto-Raman scattering}
\author{Dipankar Jana}
\email{dipankar.jana@lncmi.cnrs.fr}
\affiliation{Laboratoire National des Champs Magn\'etiques Intenses, LNCMI-EMFL, CNRS UPR3228,Univ. Grenoble Alpes, Univ. Toulouse, Univ. Toulouse 3, INSA-T, Grenoble and Toulouse, France}
\author{Piotr Kapuscinski}
\affiliation{Laboratoire National des Champs Magn\'etiques Intenses, LNCMI-EMFL, CNRS UPR3228,Univ. Grenoble Alpes, Univ. Toulouse, Univ. Toulouse 3, INSA-T, Grenoble and Toulouse, France}
\author{Amit Pawbake}
\affiliation{Laboratoire National des Champs Magn\'etiques Intenses, LNCMI-EMFL, CNRS UPR3228,Univ. Grenoble Alpes, Univ. Toulouse, Univ. Toulouse 3, INSA-T, Grenoble and Toulouse, France}
\author{Anastasios Papavasileiou}
\affiliation{Chemistry Department, University of Chemistry and Technology Prague, 16628 Prague, Czech Republic}
\author{Zdenek Sofer}
\affiliation{Chemistry Department, University of Chemistry and Technology Prague, 16628 Prague, Czech Republic}
\author{Ivan Breslavetz}
\affiliation{Laboratoire National des Champs Magn\'etiques Intenses, LNCMI-EMFL, CNRS UPR3228,Univ. Grenoble Alpes, Univ. Toulouse, Univ. Toulouse 3, INSA-T, Grenoble and Toulouse, France}
\author{Milan Orlita}
\affiliation{Laboratoire National des Champs Magn\'etiques Intenses, LNCMI-EMFL, CNRS UPR3228,Univ. Grenoble Alpes, Univ. Toulouse, Univ. Toulouse 3, INSA-T, Grenoble and Toulouse, France}
\affiliation{Institute of Physics, Charles University, Ke Karlovu 5, Prague, 121 16, Czech Republic}
\author{Marek Potemski}
\email{marek.potemski@lncmi.cnrs.fr}
\affiliation{Laboratoire National des Champs Magn\'etiques Intenses, LNCMI-EMFL, CNRS UPR3228,Univ. Grenoble Alpes, Univ. Toulouse, Univ. Toulouse 3, INSA-T, Grenoble and Toulouse, France}
\affiliation{CENTERA Labs, Institute of High Pressure Physics, PAS, 01 - 142 Warsaw, Poland}
\author{Clement Faugeras}
\email{clement.faugeras@lncmi.cnrs.fr}
\affiliation{Laboratoire National des Champs Magn\'etiques Intenses, LNCMI-EMFL, CNRS UPR3228,Univ. Grenoble Alpes, Univ. Toulouse, Univ. Toulouse 3, INSA-T, Grenoble and Toulouse, France}

\begin{abstract}
Magnon gap excitations selectively coupled to phonon modes have been studied in FePSe$_3$ layered antiferromagnet with magneto-Raman scattering experiments performed at different temperatures. The bare magnon excitation in this material has been found to be split (by $\approx~1.2$ cm$^{-1}$) into two components each being selectively coupled to one of the two degenerated, nearby phonon modes. Lifting the degeneracy of the fundamental magnon mode points out toward the biaxial character of the FePS3 antiferromagnet, with an additional in-plane anisotropy complementing much stronger, out-of-plane anisotropy. Moreover, the tunability, with temperature, of the phonon- versus the magnon-like character of the observed coupled modes has been demonstrated.

\end{abstract}

\maketitle


\section{Introduction}

van der Waals (vdW) magnets, layered materials including magnetic ions, are today at the heart of an intense effort to investigate and characterize their magnetic properties. This family of materials includes a broad variety of magnetic ground states such as ferromagnetism in CrI$_3$~\cite{Huang2017} or Cr$_2$Ge$_2$Te$_6$~\cite{Gong2017} and antiferromagnetism in the MPX$_3$ family~\cite{Jiang2021}, where M=Fe, Cr, Co, Mn, Ni and X=S or Se, together with a rich physics of interlayer ferro- or antiferromagnetic interaction across the vdW gap. Beyond the fundamental aspects related to magnetic properties of purely two-dimensional (2D) systems, the physics of 2D vdW magnets is also foreseen to play an important role in future spintronic devices~\cite{Jungwirth2016, Baltz2018} or to impose new properties on neighboring materials via proximity effects \cite{Voroshnin2022}. Magnons are among the possible elementary excitations of magnetic systems and investigating their properties provides a unique knowledge of the magnetic ground state~\cite{Lancon2016, Calder2021, Wildes2022, Wildes2023}.

The magnon-phonon interaction in magnetic materials has recently attracted a lot of interest as it determines the magnon dynamics, a property crucial for the field of magnontronics~\cite{Bozhko2020, Godejohann2020, mai2021, Liu2021, Park2016, Bao2020, Streib2019, Wang2020}. In the case of iron-based antiferromagnets, a strong single-ion anisotropy of Fe-ion boosts the magnon energy in the range of meV (with respect to $\mu eV$ for other compounds). This allows one to investigate the physics of magnon excitations using Raman scattering techniques with a typical low energy cut-off around $1$~meV and with a spatial resolution of $\sim 1 \mu m$. Such high magnon energy coincides with energies of optical phonons in covalent solids, indicating a possible influence of the magnon-phonon interaction on the observed magnon or phonon energies. Similar phenomena have been evidenced in recent works where the effects of magnon-phonon interaction are monitored by adjusting the energy detuning between the magnon and the phonon excitations by external means~\cite{Hay1969, Liu2021, Vaclavkova2021, Pawbake2022}. In the case of easy-axis antiferromagnets, up and down spins of the two  sublattices precess in the exchange field of the other one with a circular precession in opposite directions. Collective oscillations of this spin system lead to a doubly degenerate magnon excitation~\cite{RezendeBook}. In the presence of magnon-phonon interaction, as in the case of FePS$_3$, the degeneracy of the magnon gap excitation is lifted by the interaction~\cite{Vaclavkova2021, Pawbake2022}. Note that this degeneracy can also be intrinsically lifted even in the absence of magnon-phonon interaction, as it is observed in bi-axial antiferromagnets such as NiPS$_3$, MnPSe$_3$, and in CrSBr, by an additional in-plane magnetocrystalline anisotropy~\cite{Wildes2022, mai2021, Cham2022, Cho2023}. Identifying the source of magnon degeneracy is hence crucial as it brings a detailed knowledge of the magnetic system and its interacting state.

As recently reported, the magnon gap in bulk FePSe$_3$ is nearly degenerate with a pair of optical phonon modes with chiral character~\cite{Cui2023}, and the two magnon components coupled independently to the phonon with the corresponding chirality. The magnon-phonon interaction in bulk FePSe$_3$ is hence chirality selective, very distinct from the one observed in the analogous compound FePS$_3$~\cite{Liu2021, Vaclavkova2021} where both magnons couple independently to the same phonon mode. This situation makes it difficult to disentangle the effects of magnon-phonon interaction from those of magnetocrystalline anisotropy on the magnon excitation spectrum, and in particular on the degeneracy lifting at $B=0$.

In this work, we combine Raman scattering techniques with different environments of variable temperature, from $T=4.2$~K up to $T=140$~K, and of high magnetic fields up to $B=30$~T, to disentangle all coupled magnon-phonon excitations in bulk FePSe$_3$ and to pinpoint the origin of the zero-field splitting of the coupled modes in the case of selective magnon-phonon interaction. We show the presence of two non-degenerate magnon gap excitations that are selectively coupled with two  degenerate phonons, leading to the apparent splitting in the phonon modes. Our experiments indicate an intrinsic lifting of the magnon degeneracy, most probably representative of a weak magneto-crystalline anisotropy.

\begin{figure}[]
	\includegraphics[width=8.4cm]{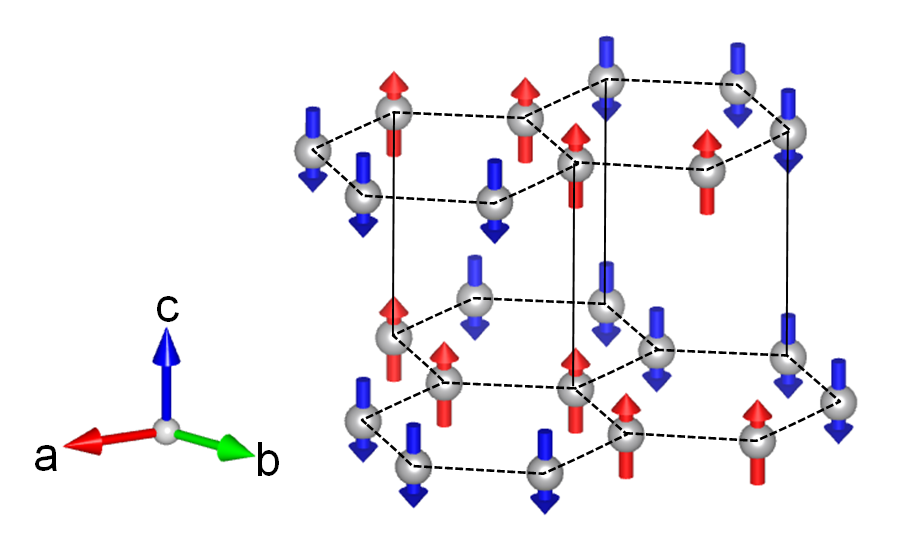}
	\caption{Magnetic structure of FePSe$_3$ in antiferromagnetic phase. The grey sphere and red/blue arrows represent Fe$^{2+}$ ion and spin direction respectively. Vertical lines guide the Fe$^{2+}$ ion staking along the c-axis. The figure is created using the VESTA software package~\cite{Momma2011}.
	\label{fig:Fig1}}
\end{figure}

\section{Materials and experimental setup}

FePSe$_3$ crystallizes in the trigonal R$\bar{3}$ space group~\cite{Wiedenmann1981, Scagliotti1987}. Within each layer, Fe atoms are surrounded by six Se atoms in octahedral coordination, while P atoms are tetrahedrally coordinated by three Se atoms and one P atom, forming a [P$_2$Se$_6$]$^{4-}$ unit. The Fe cations are organized in a honeycomb structure within the ab-plane while the layers are stacked along the c-axis. The hexagonal lattice arrangement of Fe ions of FePSe$_3$ in the antiferromagnetic phase is schematically presented in Fig.~\ref{fig:Fig1} together with the orientation of their spins. Ferromagnetic zig-zag spin chains along the b-axis are coupled antiferromagnetically with the adjacent chains along the a-axis. The spin direction is perpendicular to the layers’ plane (along c-axis). The interlayer coupling is ferromagnetic.

The investigated samples were grown by chemical vapor transport. A stoichiometric amount of Fe (99.9$\%$, -100 mesh, STREM, France), phosphorus (99.9999$\%$, granules, Wuhan Xinrong New Materials), and selenium (99.9999$\%$, granules, Wuhan Xinrong New Materials) were placed in quartz glass ampoule ($30\times180$~mm, wall thickness $3$~mm) in amount corresponding to $12$~g of FePSe$_3$ together with iodine (2mg/mL) and melt sealed under high vacuum. The ampoule was first placed in a horizontal furnace and heated at $500^{\circ}$C for 50 hours and on $600^{\circ}$C for $50$ hours. The heating and cooling rate was 1$^{\circ}$C/min. For the crystal growth, the ampoule was placed in a horizontal two-zone furnace and the first growth zone was heated on $750^{\circ}$C and the source zone was kept on $500^{\circ}$C for $2$ days, then the gradient was reversed on $700^{\circ}$C for source zone and $650^{\circ}$C for growth zone. The growth zone temperature was gradually decreased from $650^{\circ}$C to $600^{\circ}$C over a period of $10$ days.  Ampoule was cooled at room temperature and opened in an argon-filled glovebox.

Large bulk crystals were then inserted either on the cold finger of a helium flow cryostat or in a homemade setup for magneto-optical investigations. In both set-ups, the temperature can be varied from liquid helium temperature up to $T=140$~K. Magneto-optical experiments were performed in the Faraday configuration with the magnetic field applied perpendicular to the plane of the layers. The excitation laser was produced by a semiconductor-based laser ($\lambda=515$~nm) or a He-Ne gas laser ($\lambda=633$~nm), focused on the sample with a microscope objective of NA$=0.5$ (helium flow cryostat) or NA$=0.83$ (magneto-optical measurements). Scattered signals were collected using the same objective and were then analyzed by a grating spectrometer equipped with a liquid nitrogen-cooled charge-coupled device (CCD). A set of reflection-based Bragg filters are used in both the excitation and collection paths to clean the laser line and to reject the backscattered laser.

\section{Magnon-polarons in bulk $\mathrm{FePSe_{3}}$}

To unambiguously identify the magnon contribution to the global Raman scattering response, one has to change a thermodynamic parameter, such as temperature or to apply an external magnetic field, and to analyze how the different contributions to the Raman scattering spectrum evolve. The magnon energy gap at $\Gamma$, observable in Raman scattering~\cite{Fleury1968}, is expected to close due to thermal fluctuations when increasing temperature~\cite{Sekin1990} and to vanish at $T_N$.  Magnons also couple to an external magnetic field and, when increasing the magnetic field, the individual magnon components evolve in energy in a way specific to the magnetic system and to the geometry used for the magneto-Raman experiment~\cite{RezendeBook}. In the following, we will apply both approaches to identify magnon excitations, disentangle magnon-polarons and identify the source of magnon degeneracy lifting in bulk FePSe$_3$.

\begin{figure*}[]
\centering
\includegraphics[width=8.4cm]{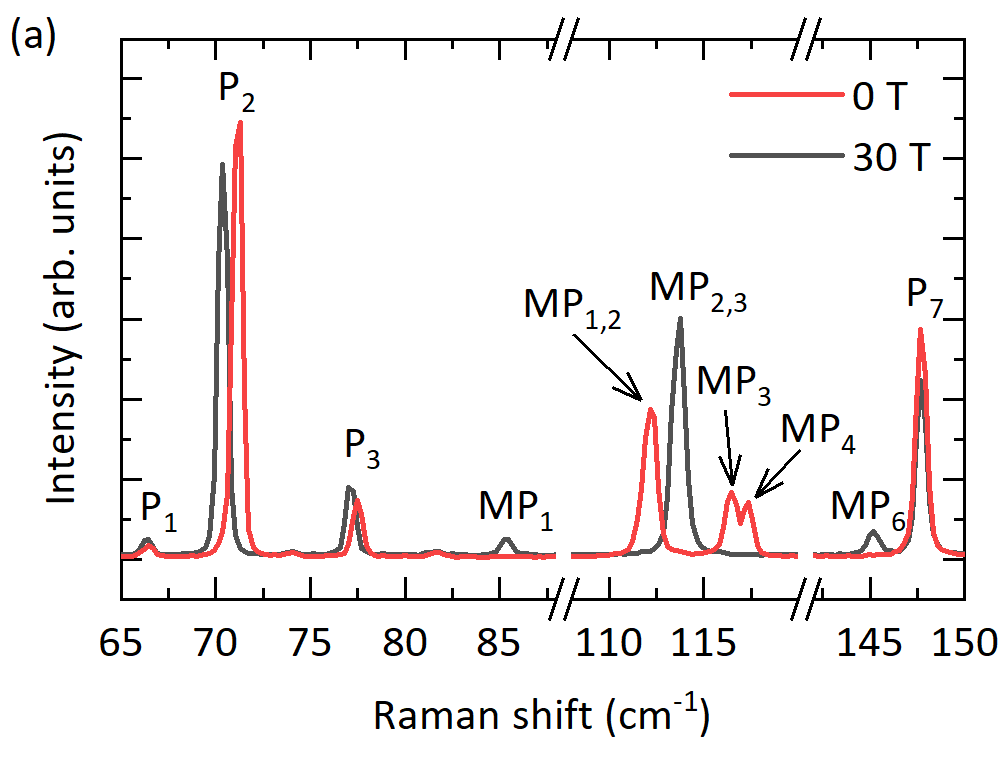}
\includegraphics[width=8.4cm]{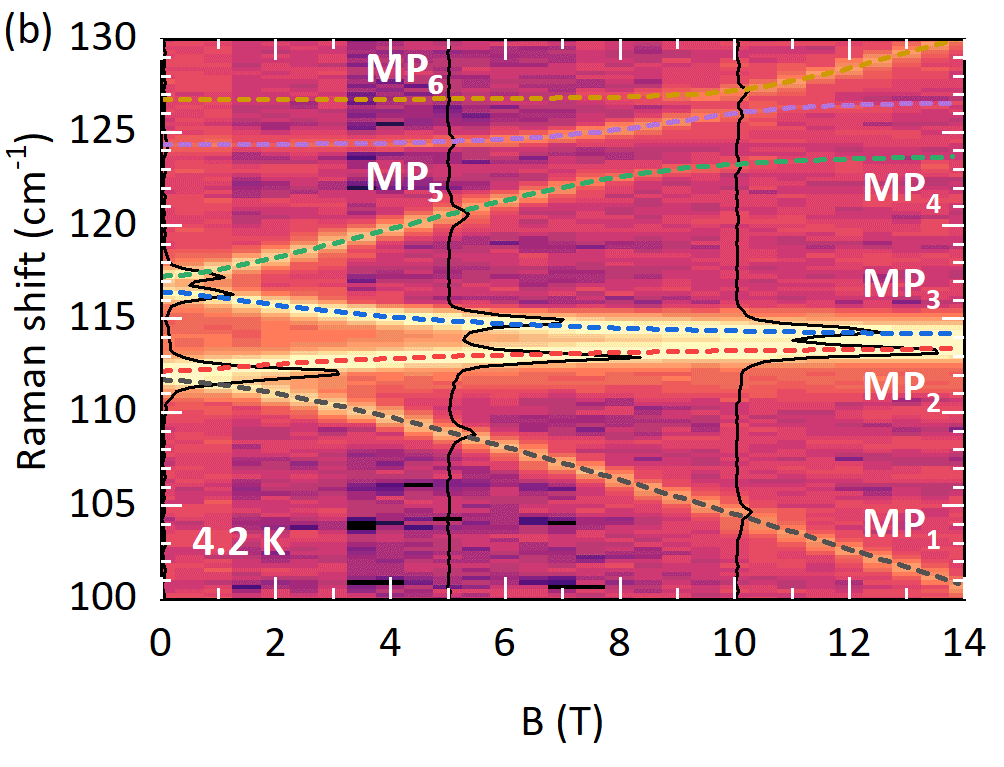}
\caption{a) $B=0$~T (red solid line) and $B=30$~T (black solid line) Raman scattering spectra of bulk FePSe$_3$ at low-temperature ($T=4.2$~K). Uncoupled phonon modes are labeled as P$_i$ and the coupled magnon-phonon modes are labeled as MP$_i$. At $B=0$~T, MP$_1$ and MP$_2$ are not resolved (see Fig.~S2 of Supplementary Materials for polarization-resolved measurements) while there is a visible splitting between MP$_3$ and MP$_4$.  (b) False color map of low-temperature Raman scattering of FePSe$_3$ as a function of the magnetic field applied along the crystal c-axis. A few representative scattering spectra, measured at magnetic field strengths of 0~T, 5~T, and 10~T are also plotted. The weak MP$_{5,6}$ peaks correspond to two additional coupled modes (not shown in Fig.~2a) which get prominent when in resonance with the high energy magnon mode. Dashed lines display magnon-polaron modes (MP$_i$) as calculated using Eq.~\ref{Ham}.
\label{fig:Fig2}}
\end{figure*}

We present in Fig.~\ref{fig:Fig2}a the unpolarized low-temperature Raman scattering response of bulk FePSe$_3$ at $B=0$~T (red curve). It is composed of several phonon modes labeled P$_1$ at $66.4$~cm$^{-1}$, P$_2$ at $71.1$~cm$^{-1}$, P$_3$ at $77.5$~cm$^{-1}$, P$_7$ at $147.7$~cm$^{-1}$, and magnon-phonon coupled modes labeled MP$_{1,2}$ at $112.2$~cm$^{-1}$, and a doublet labeled MP$_3$ and MP$_4$ at $116.4$ and $117.3$~cm$^{-1}$, respectively. The Raman scattering response of phonons of bulk FePSe$_3$ has been described previously~\cite{Scagliotti1987}. Cui et al.~\cite{Cui2023} have recently shown that MP$_{1-4}$ are coupled phonon-magnon modes, arising from a pair of quasi-degenerate chiral phonons and antiferromagnetic magnons. Interestingly the chiral phonons in bulk FePSe$_3$ are coupled selectively to the magnon component of the same chirality. In contrast to the strong coupling case, the selective coupling does not lift degeneracies either of the degenerate chiral phonons or of the degenerate AFM magnons (see also Supplementary Materials Fig. S1a and Fig. S1b-d for other possibilities). The $B=0$~T Raman scattering response of bulk FePSe$_3$ (red solid line in Fig.~\ref{fig:Fig2}a) clearly shows a splitting of the MP$_{3,4}$ peaks, while MP$_{1,2}$ appear mainly as a single peak. Using polarization-resolved low-temperature Raman scattering techniques, one can indeed resolve a splitting for both pair of modes MP$_1$/MP$_2$ ($0.2$~cm$^{-1}$) and of MP$_3$/MP$_4$ ($1.0$~cm$^{-1}$), see Fig. S2 of the Supplementary Material~\cite{SuppInfo}. The observation of an energy splitting $\Delta$MP$_{1,2}$ and $\Delta$MP$_{3,4}$ of different amplitude for both pairs of coupled modes is not consistent with the model of selective coupling among degenerate magnon and phonon, and indicates that the degeneracy of either the phonon or the magnon excitations is intrinsically lifted by a mechanism other than magnon-phonon interaction.

The application of an external magnetic field along the easy axis (c-axis) modifies the magnon excitation spectrum by inducing a linear Zeeman effect which further modifies the magnon-phonon energy detuning. The evolution of the Raman scattering response with respect to the magnetic field is presented in Fig.~\ref{fig:Fig2}b in the form of a false color plot up to $B=14$~T and the spectrum at $B=30$~T is represented in Fig.~\ref{fig:Fig2}a (black curve). While the splitting of MP$_{3,4}$ is clearly visible even at $B=0$~T, the splitting of MP$_{1,2}$  into two components with increasing magnetic field confirms the presence of two quasi-degenerate modes at $B=0$~T. The bare phonon and magnon modes are recovered with increasing magnetic field. MP$_1$ and MP$_4$ disperse linearly with the magnetic field, indicating a pronounced magnon-like character while the energy of MP$_2$ and MP$_3$ changes slightly and saturates at a constant value for $B>10$~T, revealing a phonon-like character at the high magnetic field. MP$_4$ further interacts with two more phonon modes MP$_5$ and MP$_6$ with very weak intensity, showing an avoided crossing close to $B=8$~T. Above this field, MP$_6$ displays a magnon-like behavior. As can be seen in the $B=30$~T spectrum of Fig.~\ref{fig:Fig2}a, the MP$_1$ mode is interacting with three P$_{1-3}$ low energy phonons as their energy at high field is different from that at $B=0$. Despite the close proximity with MP$_6$ at high magnetic fields, the P$_7$ phonon energy remains unaffected essentially due to the fact that this high-energy phonon does not involve the vibration in the magnetic lattice~\cite{Hashemi2017, Scagliotti1987}. An interesting observation in Fig.~\ref{fig:Fig2}a is that, at $B=30$~T (black spectrum), when the two magnon-like components MP$_{1,6}$ are well separated from each other and from the phonon modes, MP$_{2,3}$ have merged into a single peak at the bare phonon energy (See Fig.~S3 of Supplementary Materials~\cite{SuppInfo}). This is strong evidence that the bare phonon modes are degenerate. As we will demonstrate in the following, the apparent splitting between each pair of coupled modes(MP$_{1,2}$ and MP$_{3,4}$) is intrinsic to magnons in bulk FePSe$_3$ and we attribute it to a magnetocrystalline anisotropy.

\begin{figure}[]
\includegraphics{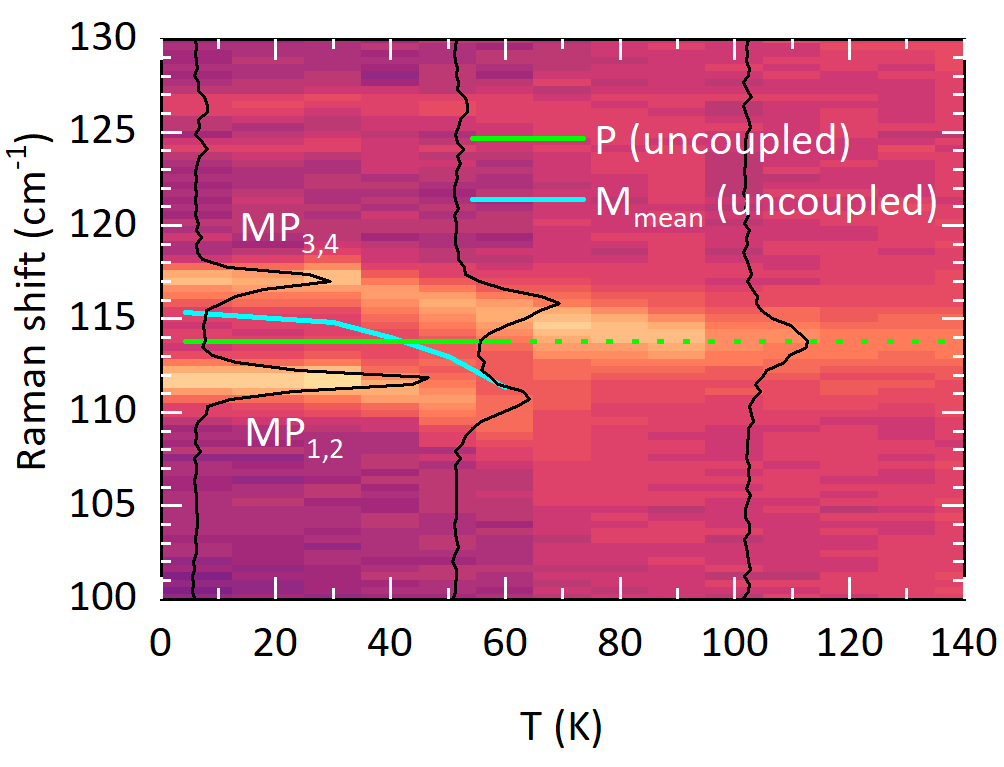}
	\caption{False color map of Raman scattering of FePSe$_3$ as a function of temperature using $\lambda=515$~nm excitation laser. Three characteristic spectra at $T=100$~K, $T=50$~K, and $T=4.2$~K are displayed. Solid lines are the bare mean magnon energy (cyan solid line) and the bare phonon energy (green solid and dotted line) as deduced from our modeling of magneto-Raman scattering experiments at different temperatures (see Fig.~\ref{fig:Fig4}a-d).
 \label{fig:Fig3}}
 \end{figure}

The energy of the coupled modes observed in Raman scattering experiments is analyzed in the frame of selective coupling between magnon and phonon as presented in Ref.~\cite{Cui2023} and considering zero-field splitting of the magnon gap excitation using the following Hamiltonian:
\begin{equation}
H_{6\times6} = \begin{bmatrix}
M_1 & 0 & 0 & \delta_1 & \delta_2 & \delta_3 \\
0 & M_2 & \delta_1 & 0 & \delta_2 & \delta_3\\
0 & \delta_1 & P_4 & 0 & 0 & 0\\
\delta_1 & 0 & 0 & P_4 & 0 & 0\\
\delta_2 & \delta_2 & 0 & 0 & P_5 & 0\\
\delta_3 & \delta_3 & 0 & 0 & 0 & P_6
\end{bmatrix}
\label{Ham}
\end{equation}
where $M_{1,2}$ are the bare energies of the two antiferromagnetic magnon branches. The magnetic field dependence of these modes is defined as $M_{1,2}(B)=M_0 \pm \sqrt{\Delta^2+(g\mu_b B)^2}$ where M$_0$ is the mean magnon energy and 2$\Delta$ the energy splitting between the magnon branches. P$_i$ are the bare phonon energies, $\delta_{1,2,3}$ are the magnon-phonon coupling parameters for P$_4$, P$_5$, and P$_6$. Each component of the doubly degenerate P$_4$ modes is coupled selectively to the two magnon components M$_1$ and M$_2$  while P$_5$ and P$_6$ couple identically to both magnon branches. We omit in this description the potential influence of $P_{1-3}$ phonons as their energy difference with the magnon is large at $B=0$~T. Using the Bogolyubov transformations\cite{white1965}, one can obtain the eigenstates of the magnon-phonon coupled modes and the corresponding coupling constants using this Hamiltonian. These results are presented in Fig.~\ref{fig:Fig2}b with dashed lines and represent the experimental evolutions accurately using the following parameters: $M_0=115.4$~cm$^{-1}$, $\Delta=0.6$~cm$^{-1}$, $g=2.14$~cm$^{-1}$, $P_4=113.8$~cm$^{-1}$, $P_5=123.9$~cm$^{-1}$, $P_6=126.7$~cm$^{-1}$ and coupling constants $\delta_1=2.36$~cm$^{-1}$, $\delta_2=1.36$~cm$^{-1}$ and $\delta_3=0.68$~cm$^{-1}$. When introducing the coupling constants, these bare modes transform into magnon-polarons MP$_i$, as labeled in Fig.~\ref{fig:Fig2} and the 2$\Delta$ energy gap is distributed among the MP$_i$ modes. The observation of larger energy splitting for MP$_{3,4}$ than for MP$_{1,2}$ (see Fig.~\ref{fig:Fig2}a) is an indication that MP$_{3,4}$ are more magnon-like while MP$_{1,2}$ are more phonon-like at $B=0$~T. This approach shows that when bulk FePSe$_3$ is magnetically ordered, the AFM magnon excitations are nearly degenerate with the pair of phonons $P_4$. This situation is seldom found in nature and, to create the magnon-polaron resonance, one usually has to tune the magnon or the phonon energies by external means.

\begin{figure*}[]
\centering
\includegraphics[width=8.4cm]{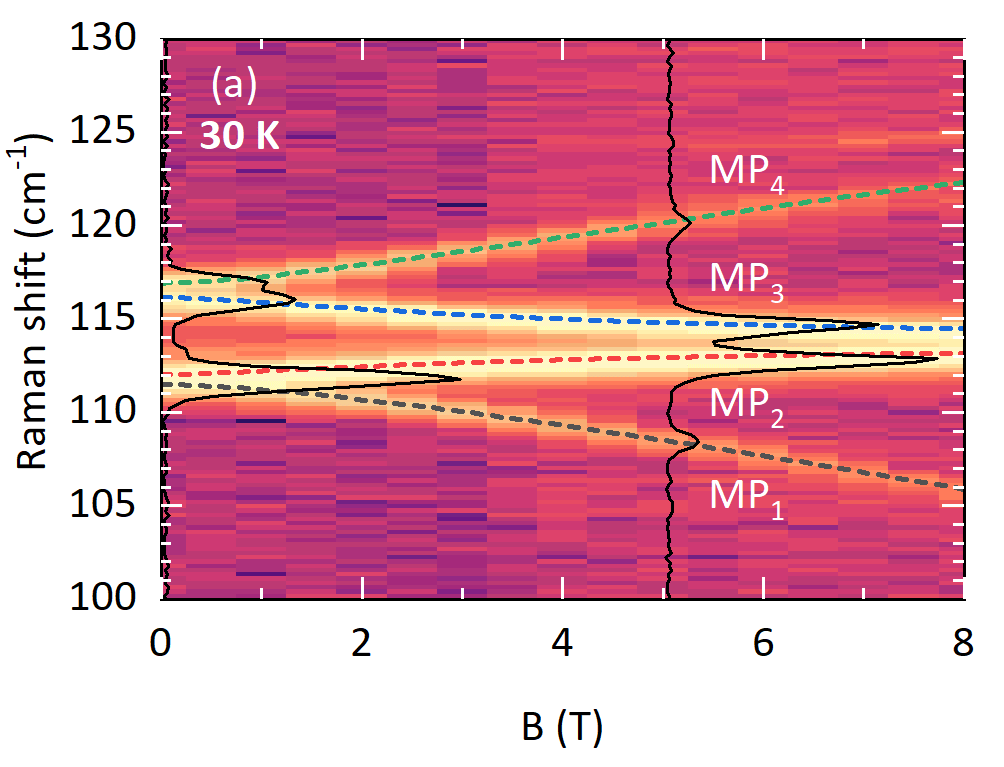}
\includegraphics[width=8.4cm]{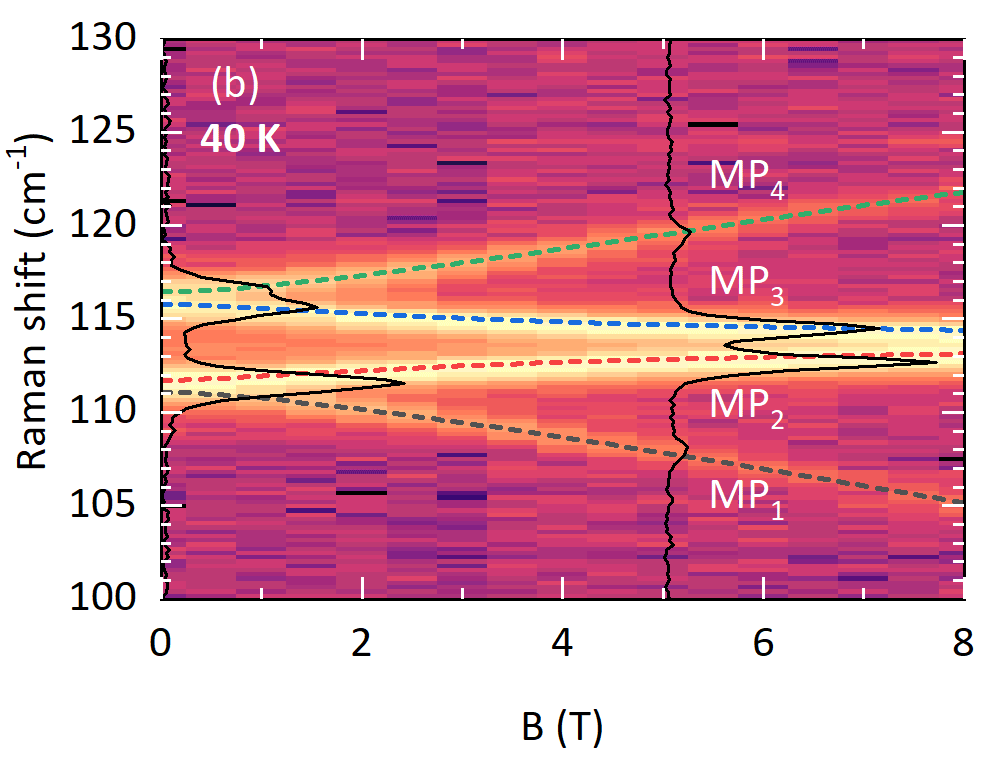}
\includegraphics[width=8.4cm]{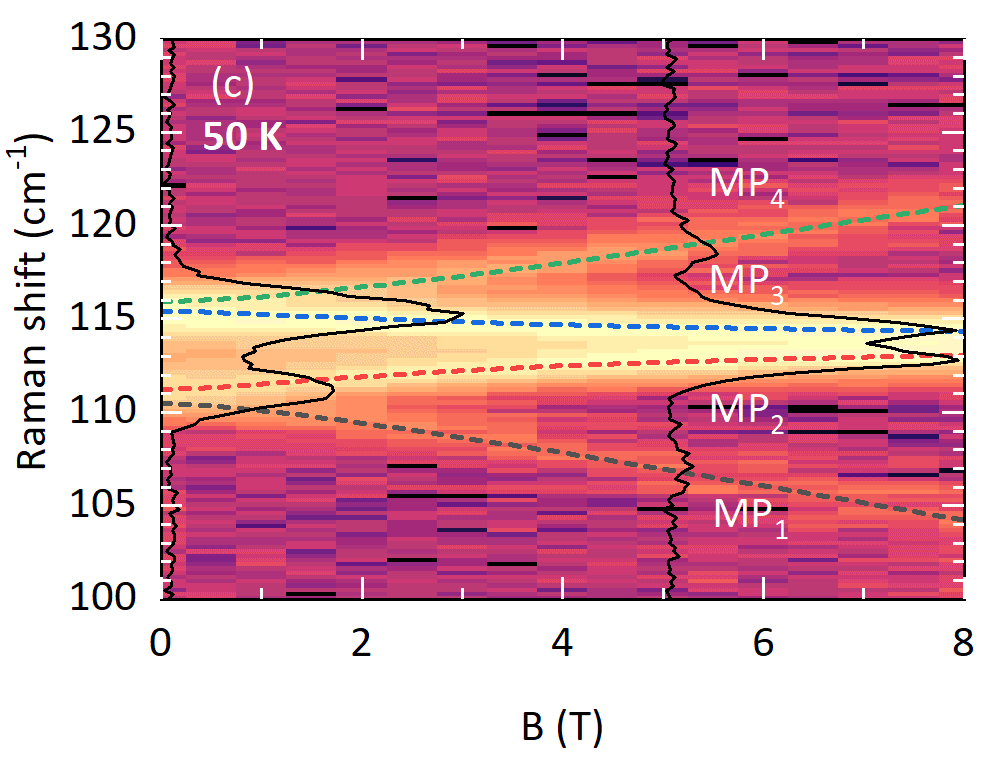}
\includegraphics[width=8.4cm]{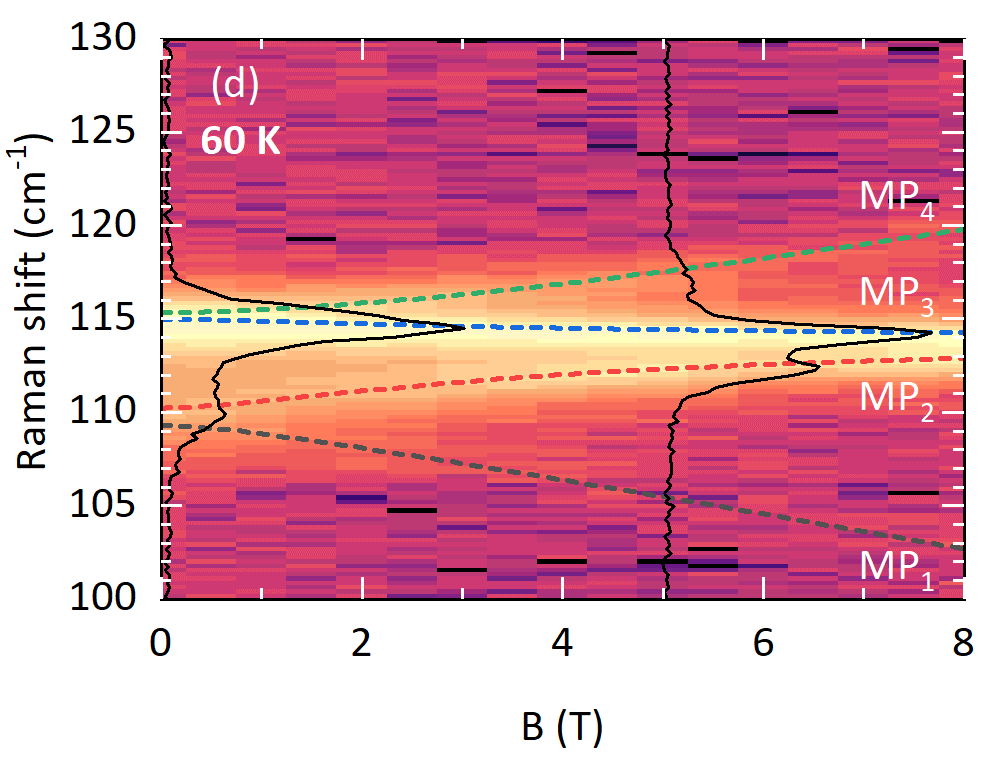}
\caption{(a-d) False color map of FePSe$_3$ Raman scattering as a function of the magnetic field applied along the crystal c-axis at temperatures 30~K, 40~K, 50~K, and 60~K respectively. Two representative scattering spectra, measured at $B=0$~T and at $B=5$~T are also plotted. Dashed lines correspond to the simulation of the magnon-polaron modes (MP$_i$) following Eq.~1.}
\label{fig:Fig4}
\end{figure*}

\begin{figure*}[]
\centering
\includegraphics[width=5.9cm]{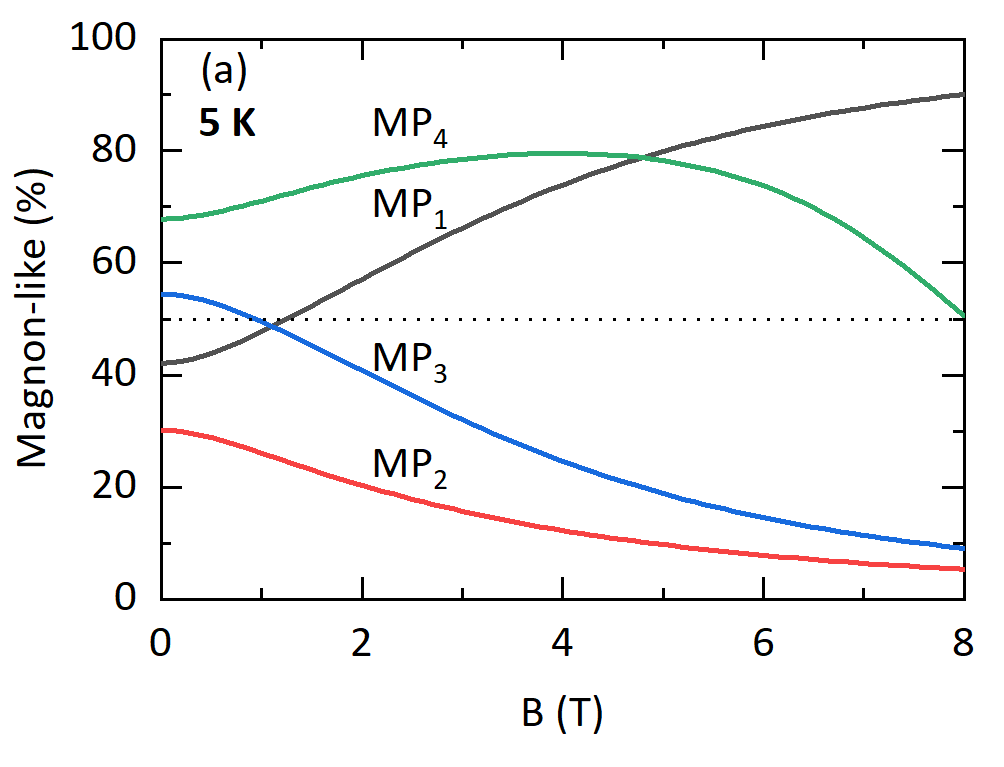}
\includegraphics[width=5.9cm]{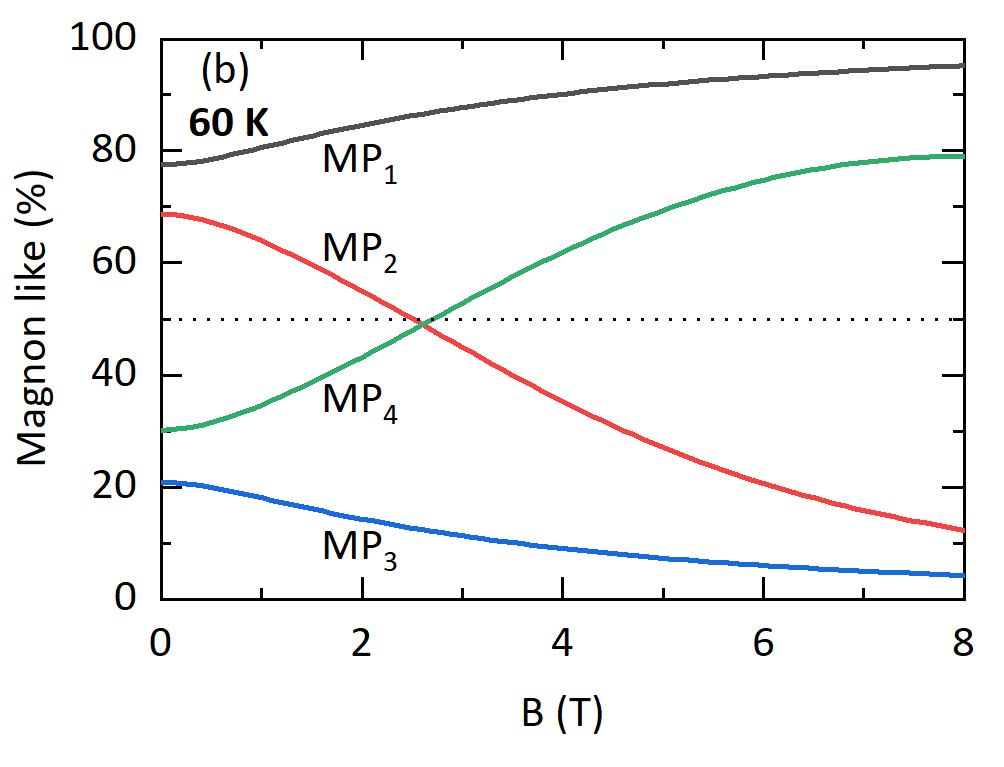}
\includegraphics[width=5.9cm]{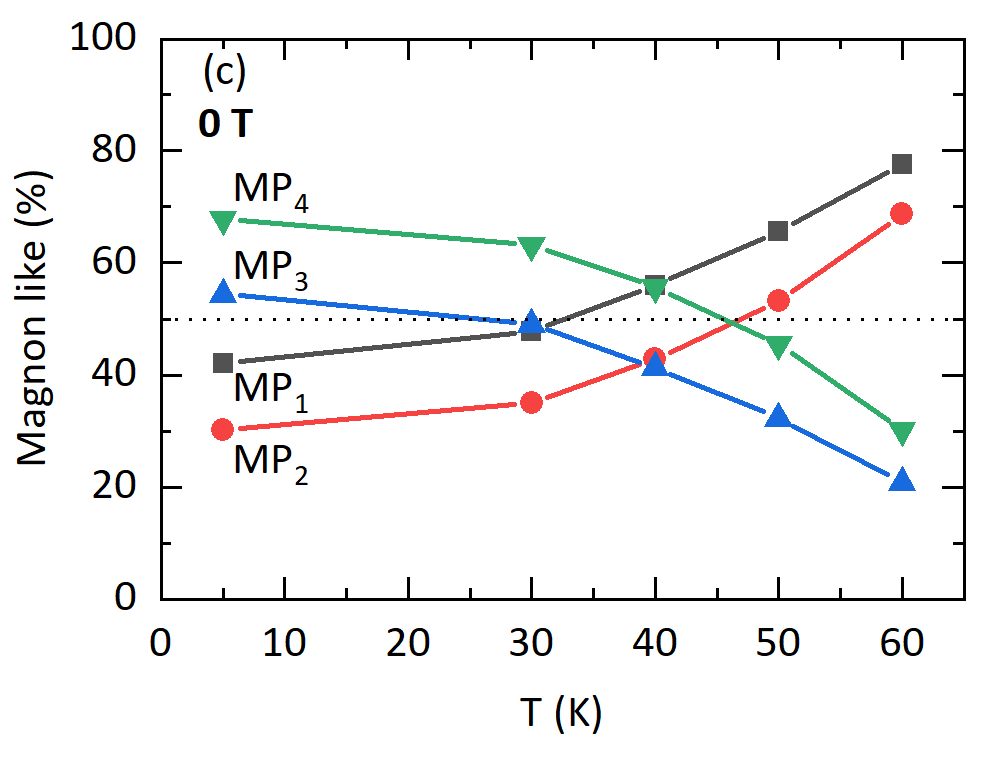}
\caption{(a) and (b) Simulation of the magnon-like behavior of the magnon-polaron modes (MP$_i$) driven by a magnetic field at 5~K and 60~K respectively. The behavior of the MP$_4$ mode is affected due to further coupling with other high-energy phonon modes (see Fig. 2a at the high field), which are not shown here. (c) Simulation of the magnon-like behavior of the magnon-polaron modes (MP$_i$) as a function of temperature at zero-field.}
\label{fig:Fig5}
\end{figure*}

Another way to control the magnon-phonon energy detuning is to change temperature. We have performed temperature-dependent Raman scattering measurements to evidence the effect of the closing of the magnon gap on magnon-polarons, see Fig.~\ref{fig:Fig3}. Starting from low temperature, magnon-polarons are first weakly affected by the change of temperature, but both MP$_{1,2}$ and MP$_{3,4}$ evolve significantly above $T=30$~K. The energy of MP$_{3,4}$ gradually changes from  $117$~cm$^{-1}$ at low temperature down to $114$~cm$^{-1}$ close to $T=90$~K and remains constant up to $T=140$~K. Simultaneously, the energy of MP$_{1,2}$ at $112$~cm$^{-1}$ also decreases along with a decrease of its intensity. Above $T=70$~K, the intensity of MP$_{1,2}$ vanishes. MP$_{1,2}$ and MP$_{3,4}$ have an avoided crossing behavior which indicates a coupling between these two pairs of modes. Though magnon excitations in this material cannot be directly observed at $T=100$~K, the emergence of the magnetic order below $T_N$ can be traced with the activation of P$_{1-3}$ phonons related to the folding of zone edges onto the $\Gamma$-point of the phonon Brillouin zone (see Fig.~S4 of Supplementary Materials~\cite{SuppInfo}).

To be able to determine accurately the parameters describing the magnetic and phonon excitations using Eq.~\ref{Ham} for different temperatures, and in particular the value of the magnon splitting energy $\Delta$, we have applied a magnetic field to separate the magnon and phonon excitations and suppress the effect of their interaction. These results are presented as false color plots of the magnetic field evolution of the Raman scattering intensity in Fig.~\ref{fig:Fig4}a-d for $T=30, 40, 50$ and $60$~K, respectively. The dashed lines in this figure are the results of the calculation using Eq.~\ref{Ham}. We do not observe significant changes in the values of the magnon g-factor, of the magnon-phonon coupling constants and the main effect of increasing temperature lies in the decrease of the bare magnon energy $M_0$. Most important is that the magnon splitting energy $\Delta$, necessary to describe Raman scattering data, appears to be independent of temperature and also of the magnon-phonon energy detuning. Because the magnon energy is decreasing when increasing temperature, the pair of MPs showing the maximum splitting at $B=0$ changes from MP$_{3,4}$ at $T=30$~K to MP$_{1,2}$ at $T=60$~K when the bare magnon energy is smaller than the bare phonon energy. This increase of the splitting of MP$_{1,2}$ with temperature is not observed in the experiment due to the thermal broadening of these Raman modes, however, MP$_{3,4}$ have merged into a single peak as shown in Fig.~S2 of the Supplementary Material~\cite{SuppInfo}, which is reproduced by our simulation. This is a strong indication that this energy splitting is intrinsic to the magnon excitation and is not imposed by the magnon-phonon interaction.

Solving Eq.~\ref{Ham} brings the coupled mode energies as eigenvalues and the eigenvectors which components describe the bare-mode content of the coupled modes together with their evolution when applying a magnetic field and changing the magnon-phonon energy detuning. The evolution of the magnon content of the MPs as a function of the magnetic field is presented in Fig.~\ref{fig:Fig5}a and b for $T=5$~K and $T=60$~K, respectively. At $T=5$~K, MP$_{3,4}$ are mainly of magnon-like character while MP$_{1,2}$ are mainly phonon-like. This character changes when the temperature is increased, as presented in Fig.~\ref{fig:Fig5}c.

When applying an external magnetic field at $T=5$~K, the magnon-like character of MP$_4$ increases significantly as the Zeeman effect increases the magnon-phonon energy detuning. Simultaneously, the magnon-like character of MP$_3$ increases. At high enough magnetic field, MP$_{1,4}$ are mostly of magnon-like character while MP$_{2,3}$ have become mostly phonon-like. At $T=60$~K, the bare magnon energy is smaller than the phonon energy and the lower-in-energy pair of MPs now shows the largest energy splitting. These MP$_{1,2}$ show the largest magnon-like character, see Fig;~\ref{fig:Fig5}b, while MP$_{3,4}$ are now mostly phonon-like. When applying a magnetic field, the magnon content of MP$_{1,4}$ increases while MP$_{2,3}$ evolves towards a phonon-like character. Extending this analysis to different temperatures, we can gain some insights into the change of the MP$_i$s content when changing the magnon-phonon energy detuning with temperature, see Fig.~\ref{fig:Fig5}c. When increasing temperature, the magnon content of MP$_{3,4}$ decreases, and they both become mainly phonon-like above $T=40$~K. MP$_{1,2}$ follows the inverse evolution, from a mainly phonon-like character at low temperature to a mainly magnon-like character above $T=40$~K. This analysis allows tracing the evolution of the bare-modes composition entering the coupled modes for different temperatures and with an applied magnetic field.

While earlier reports have considered the existence of two distinct phonon modes interacting with magnon gap excitations~\cite{Cui2023}, our experimental findings provide new evidence showing that these phonons are degenerate. Consequently, we attribute the observed zero-field splitting of both coupled modes of different amplitude, to the splitting of the underlying magnon gap excitation. One possible cause for this splitting is an additional anisotropy, induced by the substitution of S by Se atoms. Indeed, FePS$_3$ has a monoclinic structure and is an easy-axis antiferromagnet with no or very little magnetocrystalline anisotropy~\cite{Sekin1990} leading to a degenerate magnon gap excitation, while FePSe$_3$ has an orthorhombic structure and shows all signatures of an easy-axis antiferromagnet with an anisotropy that lifts the magnon gap excitation degeneracy. This situation is similar to the case MnPX$_3$ where MnPS$_3$ is an easy-axis antiferromagnet with  a monoclinic structure while MnPSe$_3$ has an orthorhombic structure and shows an additional magnetocrystalline anisotropy that lifts the degeneracy of magnon gap excitations~\cite{mai2021}.

\section{Conclusion}

Based on the Raman scattering measurements combined with an environment of variable temperature and high magnetic fields, we have evidenced selectively coupled magnon-polarons in the bulk antiferromagnet FePSe$_3$. We have used a magnon-phonon coupling matrix formalism to reproduce our experimental data and extract the bare modes energies, type of coupling, the magnon-phonon coupling constants, and the coupled mode composition in terms of bare modes. Our results clearly indicate the lifting of the degeneracy of the antiferromagnetic magnon modes (of $1.2$~cm$^{-1}$ at low temperature) in bulk FePSe$_3$ which can appear due to an additional in-plane magneto-crystalline anisotropy.

\begin{acknowledgements}

The work has been supported by the EC Graphene Flagship project. M.P. also acknowledges support from the Foundation for Polish Science (MAB/2018/9 Grant within the IRA Program financed by EU within SG OP Program). Z.S. was supported by ERC-CZ program (project LL2101) from Ministry of Education Youth and Sports (MEYS).

\end{acknowledgements}

\providecommand{\noopsort}[1]{}\providecommand{\singleletter}[1]{#1}%

\newpage
\pagenumbering{gobble}

\begin{figure}[htp]
\includegraphics[page=1,trim = 18mm 18mm 18mm 18mm,
width=1.0\textwidth,height=1.0\textheight]{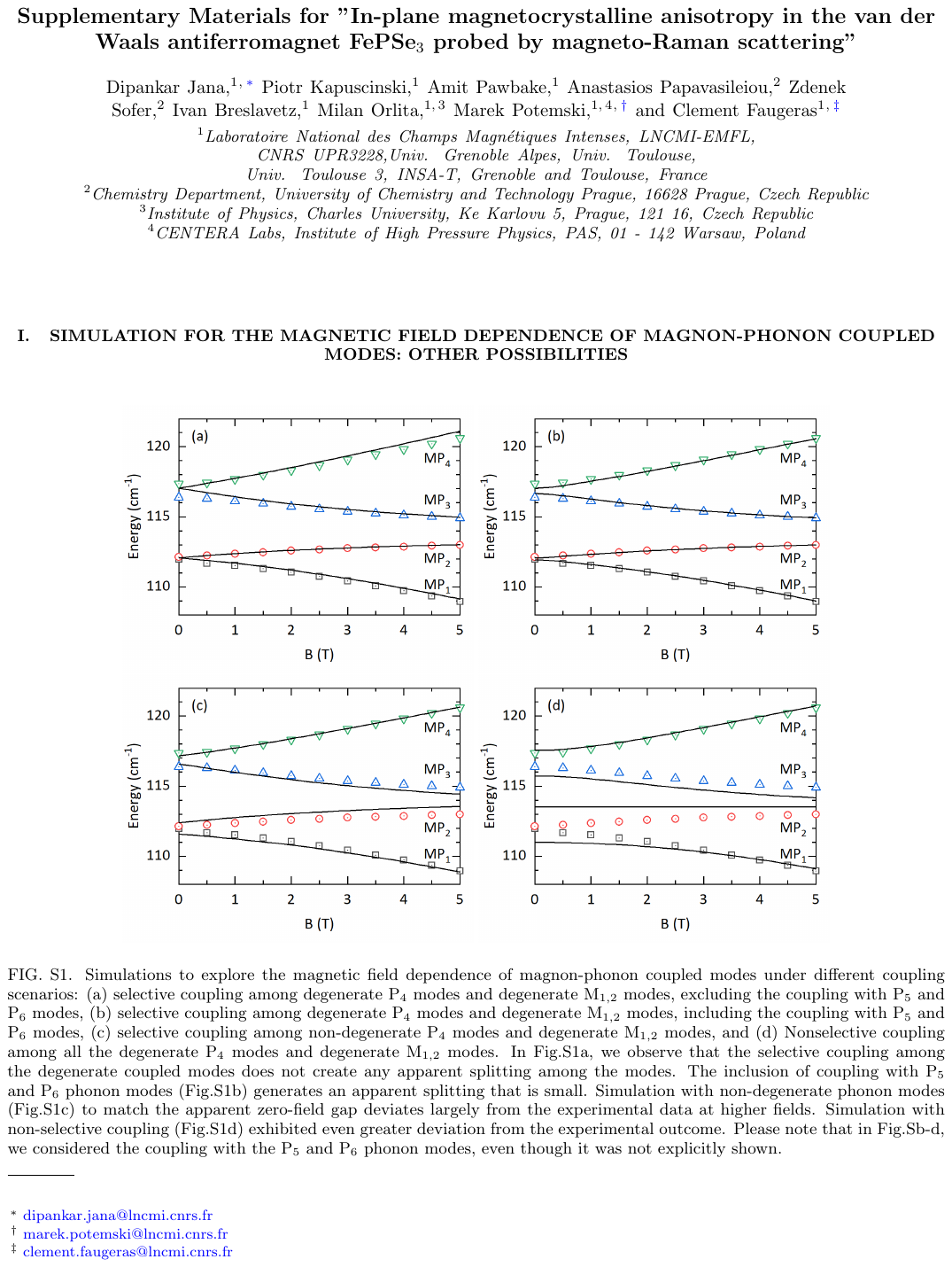}
\end{figure}

\newpage

\begin{figure}[htp]
   \includegraphics[page=2,trim = 18mm 18mm 18mm 18mm,
width=1.0\textwidth,height=1.0\textheight]{FePSe3_Magnon_polaron_SM.pdf}
\end{figure}
\newpage

\begin{figure}[htp]
   \includegraphics[page=3,trim = 18mm 18mm 18mm 18mm,
width=1.0\textwidth,height=1.0\textheight]{FePSe3_Magnon_polaron_SM.pdf}
\end{figure}

\end{document}